\documentclass[11pt]{article}
\usepackage{times}
\usepackage{fullpage}
\usepackage{epsf}
\usepackage{amsmath,amsfonts}

\usepackage{algorithmic}
\usepackage[ruled,vlined,commentsnumbered,titlenotnumbered]{algorithm2e}
\SetKwBlock{Loop}{loop}{}

\renewcommand{\th}{\ifmmode{^{\textrm{th}}}\else{\textsuperscript{th}\ }\fi}
\newcommand{\nd}{\ifmmode{^{\textrm{nd}}}\else{\textsuperscript{nd}\ }\fi}

\newcommand\tO{{\widetilde O}\/}

\newcommand\tref[1]{~\ref{#1}}
\newcommand\tsize{m}
\newcommand\cU{{\mathcal U}}
\newcommand\cR{{\mathcal R}}
\newcommand\xor{\oplus}

\newtheorem{proposition}{Proposition}
\newtheorem{lemma}[proposition]{Lemma}

\newtheorem{theorem}[proposition]{Theorem}

\newcommand{\qed}{\hbox{\rule{6pt}{6pt}}}
\newenvironment{proof}[1][]{\paragraph{Proof{#1}}}{\hfill\qed\medskip\\}

\newcommand\drop[1]{}

\newcommand{\E}{\textnormal{E}}

\def\head{{\sf head}}
\def\tail{{\sf tail}}
\newcommand\twister{h^\tau\!}
\newcommand\hT{h^\mathcal{T}\!}
\newcommand\hS{h^\mathcal{S}}

\newcommand\req[1]{(\ref{#1})}
\newcommand\eps{\varepsilon}
\newcommand\fct{\rightarrow}

\newcommand\packlist{\setlength{\itemsep}{1pt}
\setlength{\parskip}{0pt}\setlength{\parsep}{0pt}}

\newcounter{invari}
\title{Fast and Powerful Hashing using Tabulation}

\author{Mikkel Thorup\footnote{Research partly supported by Advanced
    Grant DFF-0602-02499B from the Danish Council for Independent
    Research under the Sapere Aude research carrier programme.}\\
University of Copenhagen}

\begin{document}
\maketitle
\begin{abstract}
Randomized algorithms are often enjoyed for their simplicity, but the 
hash functions employed to yield the desired probabilistic guarantees
are often too complicated to be practical. Here we survey recent results on
how simple
hashing schemes based on tabulation provide unexpectedly strong guarantees.

{\em Simple tabulation hashing\/} dates back to Zobrist [1970].  Keys are viewed as
consisting of $c$ characters and we have precomputed character tables $h_1,...,h_c$
mapping characters to random hash values. A key $x=(x_1,...,x_c)$ is hashed to
$h_1[x_1] \oplus h_2[x_2].....\oplus h_c[x_c]$. This schemes is very fast 
with character tables in cache.
While simple tabulation is not even 4-independent, it does provide
many of the guarantees that are normally obtained via higher
independence, e.g., linear probing and Cuckoo hashing.

Next we consider {\em twisted tabulation\/} where one input character
is "twisted" in a simple way. The resulting hash function has powerful
distributional properties: Chernoff-style tail bounds and a
very small bias for min-wise hashing. This also yields an extremely
fast pseudo-random number generator that is provably good for 
many classic randomized algorithms and data-structures.

Finally, we consider {\em double tabulation\/} where we compose two simple
tabulation functions, applying one to the output of the other, and
show that this yields very high independence in the classic framework of 
Carter and Wegman [1977]. In fact, w.h.p., for a given
set of size proportional to that of the space consumed, double
tabulation gives fully-random hashing. We also mention some more elaborate
tabulation schemes getting near-optimal independence for 
given time and space.

While these tabulation schemes are all easy to implement and use, their
analysis is not.
\end{abstract}

\section{Introduction}\label{sec:intro}
A useful assumption in the design of randomized algorithms and data
structures is the free availability of fully random hash functions
which can be computed in unit time. Removing this unrealistic assumption is the
subject of a large body of work. To implement a hash-based algorithm,
a concrete hash function has to be chosen. The space, time, and random
choices made by this hash function affects the overall performance.
{\em The generic goal is therefore to provide efficient constructions of hash
  functions that for important randomized algorithms yield
  probabilistic guarantees similar to those obtained assuming fully
  random hashing. }

To fully appreciate the significance of this program, we note that many
randomized algorithms are very simple and popular in practice, but often
they are implemented with too simple hash functions without the necessary
guarantees. This may work very well in random tests, adding to their
popularity, but the real world is full of structured data, e.g., generated
by computers, that could be bad for the hash function.
This was illustrated in \cite{thorup12kwise}
showing how simple common inputs made linear probing fail with popular
hash functions, explaining its perceived unreliability in practice. The
problems disappeared when sufficiently strong hash functions were used.

In this paper we will survey recent results from
\cite{CPT15:indep,DKRT15:k-part,DKRT16:two-choice,DT14:twist-min,patrascu12charhash,PT13:twist,Tho13:simple-simple}
showing how simple realistic hashing schemes based on tabulation
provide unexpectedly strong guarantees for many popular randomized
algorithms, e.g., linear probing,
Cuckoo hashing, min-wise independence, treaps, planar partitions, 
power-of-two-choices, Chernoff-style concentration bounds, and even
high independence. The survey is from a users perspective,
explaining how these tabulation schemes can be applied. While these
schemes are all very simple to describe and use, the analysis showing
that they work is non-trivial. For the analysis,
the reader will be referred to the above papers. The reader is also
referred to these papers for a historical account of previous work.

\paragraph{Background}
Generally a hash function maps a key universe $\cU$ of keys into some range
$\cR$ of hash values. A random hash function $h$ is a random variable from
$\cR^\cU$, assigning a random hash value $h(x)\in\cR$ to every $x\in\cU$.
A truly random hash function is picked uniformly from $\cR^\cU$,
assigning a uniform and independent hash value $h(x)\in \cR$ to
  each key $x\in \cU$.
Often randomized algorithms are analyzed assuming access to truly
random hash functions. However, just storing a truly random hash
function requires $|\cU|\log_2 |\cR|$ bits, which is unrealistic for large
key universes.

In general, the keys may originate come from a very large universe
$\cU$. However,
often we are only interested in the performance on an unknown
set $S\subseteq \cU$ of up to $n$ keys. Then our first step is to do a {\em universe
reduction}, mapping $\cU$ randomly to ``signatures'' in $[u]=\{0,1,\ldots,u-1\}$, 
where $u=n^{O(1)}$, e.g., $u=n^3$, so that no two keys from $S$ are expected to
get the same signature \cite{carter77universal}. Below we generally assume that this
universe reduction has been done, if needed, hence that we ``only'' need to
deal with keys from the polynomial universe $[u]$.

The concept of $k$-independence was introduced by Wegman and
Carter~\cite{wegman81kwise} in FOCS'79 and has been the cornerstone of
our understanding of hash functions ever since. As above, we think of a hash function $h: [u] \to [\tsize]$ as a random variable distributed over
$[\tsize]^{[u]}$. We say that $h$ is $k$-independent
if (a) for any distinct keys $x_1, \dots, x_k \in [u]$, the hash values
$h(x_1), \dots, h(x_k)$ are independent random variables; and (b) for
any fixed $x$, $h(x)$ is uniformly distributed in $[\tsize]$.

As the concept of independence is fundamental to probabilistic
analysis, $k$-independent hash functions are both natural and powerful in
algorithm analysis. They allow us to replace the heuristic assumption
of truly random hash functions that are uniformly distributed in $[\tsize]^{[u]}$, hence
needing $u\lg \tsize$ random bits ($\lg=\log_2$), with real implementable hash
functions that are still ``independent enough'' to yield provable
performance guarantees similar to those proved with true randomness. We are then left with the natural goal of
understanding the independence required by hashing-based algorithms. 

Once we have proved that $k$-independence suffices for a hashing-based 
randomized algorithm, we are free to use {\em any\/} $k$-independent hash function.
The canonical construction of a $k$-independent hash function is based on
polynomials of degree $k-1$. Let $p \ge u$ be prime. Picking random
$a_0, \dots, a_{k-1} \in \{0, \dots, p-1\}$, the hash function is
defined by:
\begin{equation}\label{eq:poly}
h(x) = \Big( \big( a_{k-1} x^{k-1} + \cdots + a_1 x + a_0 \big)
        \bmod{p} \Big)
\end{equation}
If we want to limit the range of hash values to $[m]$, we use
$h(x)\bmod m$. This preserves requirement (a) of independence among
$k$ hash values. Requirement (b) of uniformity is close to 
satisfied if $p\gg m$. As suggested in  \cite{carter77universal},
for a faster implementation, we can let $p$ be a Mersenne prime, e.g.,
to hash 64-bit integers, we could pick $p=2^{81}-1$.

Sometimes 2-independence suffices. For example, 2-independence implies
so-called universality \cite{carter77universal}; namely that the probability
of two keys $x$ and $y$ colliding with $h(x)=h(y)$ is $1/\tsize$; or close
to $1/m$ if the uniformity of (b) is only approximate.
Universality implies expected constant time performance of hash tables
implemented with chaining. Universality also suffices
for the 2-level hashing of Fredman et al.
\cite{fredman84dict}, yielding static hash tables with constant query time.
Moreover, Mitzenmacher and Vadhan \cite{mitzenmacher08hash} have proved
that 2-independent hashing in many applications works almost like truly random hashing if the
input has enough entropy. However, structured, low-entropy data, are very common in the real world.

We do have very fast implementations of universal 
and 2-independent hashing  \cite{dietzfel96universal,dietzfel97closest}, but unfortunately, these methods do not generalize nicely to higher independence.

At the other end of the spectrum, when dealing with problems involving $n$ objects, $O(\lg n)$-independence suffices in
a vast majority of applications. One reason for this is the Chernoff
bounds of~\cite{schmidt95chernoff} for $k$-independent events, whose
probability bounds differ from the full-independence Chernoff bound by
$2^{-\Omega(k)}$. Another reason is that random graphs with $O(\lg
n)$-independent edges~\cite{alon08kwise} share many of the properties
of truly random graphs.

When it comes to high independence, we note that the polynomial method
from Equation \req{eq:poly} takes $O(k)$ time  and space for $k$-independence. This is
no coincidence. Siegel \cite{siegel04hash}
has proved that to implement $k$-independence
with less than $k$ memory accesses, we need a representation using
$u^{1/k}$ space. He also gives a solution that for any $c$ uses
$O(u^{1/c})$ space, $c^{O(c)}$ evaluation time, and achieves
$u^{\Omega(1/c^2)}$ independence (which is superlogarithmic, at least
asymptotically, for $c=O(1)$). The construction is non-uniform, assuming a certain
small expander which gets used in a graph product. Siegel \cite{siegel04hash}
states about his scheme that it is 
``far too slow for any practical application''.

The independence measure has long been central to the study of
randomized algorithms. For example, \cite{karloff93prg} considers variants of
QuickSort, \cite{alon99linear} consider the maximal bucket size for
hashing with chaining, and   
\cite{cohen09cuckoo5,dietzfelbinger09cuckoo-bas} consider Cuckoo hashing.
In several cases \cite{alon99linear,dietzfelbinger09cuckoo-bas,karloff93prg},
it is proved that linear transformations $x\mapsto \big( (ax + b) \bmod p \big)$ do not suffice for good performance, hence that 
2-independence is not in itself sufficient.

This paper surveys a family of ``tabulation'' based hash function
that like Siegel's hash function use $O(u^{1/c})$ space for
some constant $c$, but which are simple
and practical, and offer strong probabilistic guarantees for many popular
randomized algorithms despite having low independence. We start with
the simplest and fastest tabulation scheme, and move later to more complicated schemes with stronger probabilistic guarantees.

We note that there has been several previous works using tabulation to
construct efficient hash functions (see, e.g., \cite{dietzfel90highperf,dietzfel03tabhash}), but the
schemes discussed here are simpler and more efficient. More detailed
comparisons with previous works are found in the papers surveyed.

\section{Simple tabulation}\label{sec:simple-intro}
The first scheme we consider is {\em simple tabulation\/} hashing where
the hash values are $r$-bit numbers. Our goal is to hash keys
from $\cU=[u]$  into the range $\cR = [2^r]$. In tabulation hashing, a key
$x\in[u]$ is interpreted as a vector of $c>1$
``\emph{characters}'' from the alphabet $\Sigma=[u^{1/c}]$, i.e.,
$x=(x_0,...,x_{c-1})\in\Sigma^c$. As a slight abuse of notation, we
shall sometimes use $\Sigma$ instead of $|\Sigma|$ to denote the size
of the alphabet when the context makes this meaning clear. This
matches the classic recursive set-theoretic definition of a natural as
the set of smaller naturals.

For ``simple tabulation hashing'' we initialize independent
random character tables $h_0, \dots, h_{c-1}:\Sigma\rightarrow \cR$.
The hash $h(x)$ of a key $x=(x_0,...,x_{c-1})$ is computed as:
\begin{equation}\label{eq:simple-table} 
h(x) = \bigoplus_{i\in[c]} h_i[x_i].
\end{equation}
Here $\xor$ denotes bit-wise exclusive-or. This is a well-known
scheme dating back at least to Zobrist
\cite{zobrist70hashing}. For him a character position corresponds
to a position on a game board, and the character is the piece at the
position. If the piece at a position $i$ changes from $x_i$ to $x_i'$, 
he updates the overall hash value $h$ to $h'=h\xor h_i[x_i]\xor h_i[x_i']$.

It is easy to see that simple tabulation is 3-independent, for if we
have a set $X$ of two or three keys, then there must be a position
$i\in [c]$ where one key $x$ has a character $x_i$ not shared with any other
key in $X$. This means that $x$ is the only key in
$X$ whose hash value depends on $h_i[x_i]$, so the hash value of $x$
is independent of the other hash values from $X$. On the other
hand, simple tabulation is not $4$-independent. Given $4$ keys $(a_0,b_0)$,
$(a_1,b_0)$, $(a_0,b_1)$, $(a_1,b_1)$, no matter how we fill our
tables, we have
\[h(a_0,b_0)\xor h(a_1,b_0)\xor h(a_0,b_1) \xor h(a_1,b_1)=0.\]
Thus, given the hash values of any of three of the keys, we can uniquely determine the fourth hash value.

In our context, we assume that the number $c=O(1)$ of character positions is
constant, and that character tables fit in fast cache. Justifying this
assumption, recall that if we have $n$ keys from a very large universe, we can
first do a universe reduction, applying a universal hashing function \cite{carter77universal} into an
intermediate universe of size $u=n^{O(1)}$, expecting no collisions
(sometimes, as in \cite{thorup09linprobe},  we can even accept some collisions from the first universal
hashing, and hence use an intermediate universe size closer to
$n$). Now, for any desired small
constant $\eps>0$, we get down to space $O(n^{\eps})$ picking
$c=O(1)$ so that $\Sigma=u^{1/c}<n^{\eps}$.
We shall refer to the lookups in the $h_i$ as ``character lookups'',
emphasizing that they are expected to be much faster than a general
lookup in a table of size $n$.

Putting things into a practical perspective (this paper claims 
practical schemes with strong theoretical guarantees), in the
experiments from \cite{patrascu12charhash,thorup11timerev}, for 32-bit
keys, simple tabulation with 4 character lookups took less than 5ns
whereas a single memory lookup in a 4MB table took more than
120ns. Character lookups were thus about 100 times faster than general
lookups! We note that the character lookups parallelize easily and
this may or may not have been exploited in the execution (we just wrote portable
C-code, leaving the rest to compiler and computer).  Simple tabulation
was only 60\% slower than the 2-independent multiply-shift scheme from
\cite{dietzfel96universal} which for 32-bit keys is dominated by a
single 64-bit multiplication. However, simple tabulation is
3-independent and in experiments from \cite{patrascu12charhash}, it
was found to
be more than three times faster than 3-independent hashing implemented 
by a
degree-2 polynomial tuned over the Mersenne prime field $\mathbb
Z_{2^{61}-1}$. The high speed of simple tabulation conforms with
experiments on much older architectures
\cite{thorup00universal}. Because cache is so critical to computation,
most computers are configured with a very fast cache, and this is
unlikely to change.

Usually it is not a problem to fill the character tables $h_0, \dots,
h_{c-1}$ with random numbers, e.g., downloading them from
\texttt{http://random.org} which is based on atmospheric noise. However, for the
theory presented here, it would suffice to fill them with a strong
enough pseudo-random number generator (PRG), like a $(\lg
u)$-independent hash function, e.g., using the new fast generation from \cite{christiani14prg}. The character tables just need to point to an
area in memory with random bits, and this could be shared across many
applications. One could even imagine computers configured with 
random bits in some very fast read-only memory allowing parallel access
from multiple cores.

In \cite{patrascu12charhash} simple tabulation hashing was proved to
have much more power than suggested by its 3-independence. This included
fourth moment bounds, min-wise hashing, random graph properties
necessary in cuckoo hashing \cite{pagh04cuckoo}, and Chernoff bounds for distributing
balls into {\em many\/} bins. The details follow below.

\paragraph{Concentration bounds}
First we consider using simple tabulation hashing to distribute $n$
balls into $m=2^r$ bins, that is, assuming that the balls have keys
from $[u]$, we are using a simple tabulation
hash function $h:[u]\rightarrow[m]$. In a hash table
with chaining, the balls in a bin would be stored in a linked list.

Consider the number $X$ of balls landing in a given bin. We have
$\mu=\E[X]=n/m$.  P\v{a}tra\c{s}cu and Thorup \cite{patrascu12charhash} have
proved that, w.h.p., we get a Chernoff-style concentration on $X$. 
First recall the classic Chernoff bounds~\cite[\S 4]{motwani95book} for
full randomness. On the upper bound side, we have~\cite[Theorem 4.1]{motwani95book} 
\begin{equation}\label{eq:classic-chernoff+}
\Pr[X\ge (1+\delta)\mu] \le \left(
   \frac{e^\delta}{(1+\delta)^{(1+\delta)}} \right)^{\mu} \left[\;\leq\ 
\exp(-\delta^2\mu/3)\textnormal{ for }\delta\leq 1\right]
\end{equation}
The corresponding probabilistic lower bound~\cite[Proof of Theorem
  4.2]{motwani95book} for $\delta \leq 1$ is
\begin{equation}\label{eq:classic-chernoff-}
\Pr[X\le (1-\delta)\mu] \le \left(
   \frac{e^{-\delta}} {(1-\delta)^{(1-\delta)}} \right)^{\mu}
 \left[\;\leq\ 
\exp(-\delta^2\mu/2)\textnormal{ for }\delta\leq 1\right]
\end{equation}
We note that in connection with hash tables, we are often not just
interested in a given bin, but rather we care about the bin that a
specific query ball lands in. This is why the hash of the query
ball is involved in the theorem below with Chernoff-style bounds for simple tabulation. 
\begin{theorem}[{\cite{patrascu12charhash}}]  \label{thm:chernoff} 
Consider hashing $n$ balls into $m\ge n^{1-1/(2c)}$ bins by simple tabulation (recall that $c=O(1)$ is the 
number of characters). Define $X$ as the number of regular balls that
hash into a given bin or a bin chosen as a function of the bin $h(q)$
of an additional \emph{query ball} $q$. The following probability
bounds hold for any constant $\gamma$:
\begin{align}\label{eq:char+}
\Pr[X\ge (1+\delta)\mu] &\le \left(
   \frac{e^\delta}{(1+\delta)^{(1+\delta)}} 
             \right)^{\Omega\drop{_{\gamma,c, \eps}}(\mu)} + 
    1/m^\gamma\\
\Pr[X\le (1-\delta)\mu] &\le \left(
   \frac{e^{-\delta}} {(1-\delta)^{(1-\delta)}} \right)^{\Omega\drop{_{\gamma,
       c, \eps}}(\mu)} + 1/m^\gamma.\label{eq:char-}
\end{align}
With $m\leq n$ bins (including $m< n^{1-1/(2c)}$), every bin gets
\begin{equation}\label{eq:few-bins}
n/m\pm O\left(\sqrt{n/m}\log^c n\right).
\end{equation}
balls with probability $1-n^{-\gamma}$.
\end{theorem}

Contrasting the standard Chernoff bounds, we see that 
Theorem \ref{thm:chernoff} \req{eq:char+} and
\req{eq:char-} can only provide polynomially small probability, i.e.~at
least $m^{-\gamma}$ for any desired constant $\gamma$. This corresponds
to if we had $\Theta(\log m)$-independence in
the Chernoff bound from~\cite{schmidt95chernoff}. In addition,
the exponential dependence on $\mu$  is reduced by a constant which depends
(exponentially) on the constants $\gamma$ and $c$. 

The upper bound \req{eq:char+} implies
that any given bin has 
$O(\lg n / \lg\lg n)$ balls w.h.p., but then this holds for all $m$ bins w.h.p.  Simple tabulation is the simplest and fastest constant time
hash function to achieve this fundamental property. 

Complementing the above Chernoff-style bound, Dahlgaard et
al. \cite{DKRT15:k-part} have proved that we also 
get the $k\th$ moment bounds normally associated with $k$-independence.
\begin{theorem}[{\cite{DKRT15:k-part}}]   \label{thm:k-moment} With the same setup as in Theorem \ref{thm:chernoff}, for any constant $k=O(1)$,
\[\E\left[(X-\mu)^k\right]=O(\mu+\mu^{k/2}).\]
\end{theorem}
For $k\leq 4$, this bound was proved in \cite{patrascu12charhash}, and
disallowing the dependence on a query ball, it was also proved in
\cite{braverman10kwise}, again for $k\leq 4$. 
Compelling
applications of 4\th moment bounds are given by   \cite{alon96frequency},
\cite{karloff93prg},
and \cite{pagh07linprobe}. In \cite{karloff93prg}, it was shown that
any hash function with a good 4\th moment bound suffices for a
non-recursive version of QuickSort, routing on the hypercube, etc. 
In   \cite{alon96frequency}, the 4\th moment is used to 
estimate the 2\nd moment of a data stream. In \cite{pagh07linprobe},
limited 4\th moment is shown to imply constant expected performance for
linear probing. The applications in \cite{karloff93prg,pagh07linprobe}
both require dependence of a query bin.

Since $k\th$ moment bounds is one of the main ways $k$-independence is used,
it is nice that they are achieved by simple tabulation which is
only 3-independent.

\paragraph{Linear probing}
Theorem \ref{thm:chernoff} is used in \cite{patrascu12charhash}
to get bounds for linear probing. Linear probing is a classic implementation of hash tables. It uses a
hash function $h$ to map a set of $n$ keys into an array of size
$\tsize$. When inserting $x$, if the desired location $h(x)\in
[\tsize]$ is already occupied, the algorithm scans $h(x)+1, h(x)+2,
\dots,\tsize-1,0,1,\ldots$ until an empty location is found, and
places $x$ there. The query algorithm starts at $h(x)$ and scans
until it either finds $x$, or runs into an empty position, which
certifies that $x$ is not in the hash table.  When the query search is
unsuccessful, that is, when $x$ is not stored, the query algorithm
scans exactly the same locations as an insert of $x$. A general bound
on the query time is hence also a bound on the insertion time.

This classic data structure is one of the most popular implementations of hash
tables, due to its unmatched simplicity and efficiency. 
The practical use of linear probing dates back at least to 1954 to an
assembly program by Samuel, Amdahl, Boehme (c.f. \cite{knuth-vol3}).
On modern
architectures, access to memory is done in cache lines (of much more
than a word), so inspecting a few consecutive values typically
translates into a single memory access. Even if the scan straddles a
cache line, the behavior will still be better than a second random
memory access on architectures with prefetching. Empirical
evaluations~\cite{black98linprobe,heileman05linprobe,pagh04cuckoo}
confirm the practical advantage of linear probing over other known
schemes, while cautioning~\cite{heileman05linprobe,thorup12kwise} that
it behaves quite unreliably with weak hash functions.

Linear probing was shown to take expected constant time for
any operation in 1963 by Knuth~\cite{knuth63linprobe}, in a report which is now
regarded as the birth of algorithm analysis. This analysis, however, 
assumed a truly random hash function. However, 
Pagh et al.~\cite{pagh07linprobe} showed that just $5$-independence
suffices for this expected constant operation time. In 
\cite{patrascu10kwise-lb}, $5$-independence was proved to be
best possible, presented a concrete combination of keys and a 4-independent random hash function
where searching certain keys takes $\Omega(\log n)$ expected time. 

In \cite{patrascu12charhash}, the result from~\cite{pagh07linprobe} is
strengthened for more filled linear probing tables, showing
that if the  table
size is $m=(1+\eps) n$, then the expected time per operation is
$O(1/\eps^2)$, which asymptotically matches the bound of 
Knuth \cite{knuth63linprobe} with truly random hashing. More important for this paper, \cite{patrascu12charhash}
proved that this performance bound also holds with simple tabulation hashing.

In fact, for simple tabulation, we get quite strong concentration results
for the time per operation, e.g,, constant variance
for constant $\eps$. For contrast, with 5-independent hashing,
the variance is only known to be $O(\log n)$
\cite{pagh07linprobe,thorup12kwise}. 

Some experiments are done in \cite{patrascu12charhash} comparing simple
tabulation with the fast 2-independent multiply-shift scheme from
\cite{dietzfel96universal} in linear probing. For simple inputs such as consecutive integers, the performance was extremely
unreliable with the 2-independent hashing, but with simple tabulation,
everything worked perfectly as expected from the theoretical guarantees.

\paragraph{Cuckoo hashing}
In cuckoo hashing \cite{pagh04cuckoo} we use two tables of size $m \ge
(1+\eps) n$ and independent hash functions $h_0$ and $h_1$ mapping the
keys to these two tables. Cuckoo hashing succeeds if we can place
every key in one of its two hash locations without any collision. We
can think of this as a bipartite graph with a set for each table and
an edge $(h_0(x),h_1(x))$ for each key $x$. Cuckoo hashing fails
exactly if this graph has a component with more edges than vertices.
With truly random hashing, this bad event happens with probability
$\Theta(\frac{1}{n})$. P\v{a}tra\c{s}cu and Thorup \cite{patrascu12charhash} study the random graphs induced by
simple tabulation, and obtain a rather unintuitive result: the worst
failure probability is inversely proportional to the \emph{cube root}
of the set size.

\begin{theorem}[{\cite{patrascu12charhash}}] \label{thm:cuckoo}
Any set of $n$ keys can be placed in two tables of size $m=(1+\eps)$ by
cuckoo hashing and simple tabulation with probability $1 -
O(n^{-1/3})$. There exist sets on which the failure probability is
$\Omega(n^{-1/3})$.
\end{theorem}

Thus, cuckoo hashing with simple tabulation is an excellent
construction for a static dictionary. The dictionary can be built (in
linear time) after trying $O(1)$ independent hash functions w.h.p.,
and later every query runs in constant worst-case time with two
probes. We note that even though cuckoo hashing requires two
independent hash functions, these essentially come for the cost of one
in simple tabulation: the pair of hash codes can be stored
consecutively, in the same cache line, making the running time
comparable with evaluating just one hash function.

In the dynamic case, Theorem \ref{thm:cuckoo} implies that 
we expect $\Omega(n^{4/3})$ updates between failures
requiring a complete rehash with new hash functions.

\paragraph{Minwise independence}
In \cite{patrascu12charhash} it is shown that simple tabulation is
$\eps$-minwise independent, for a vanishingly small $\eps$ (inversely
polynomial in the set size $n$).  This takes $\Theta(\log
n)$-independence by general techniques
\cite{indyk01minwise,patrascu10kwise-lb}. More precisely, we have

\begin{theorem}[{\cite{patrascu12charhash}}]\label{thm:char-min}
Consider a set $S\subseteq \Sigma^c$ of $n=|S|$ keys and $q\in S$. If
$h:\Sigma^c\rightarrow [m]$, $m\geq n^{1+1/c}$ is implemented by simple tabulation:
\begin{equation}\label{eq:char-min} \Pr[ h(q) = \min h(S)] = \frac{1\pm \eps}{n}, \qquad
\textrm{where } \eps = O\left( \frac{\lg^2 n}{n^{1/c}} \right). 
\end{equation}
\end{theorem}

The classic application of $\eps$-minwise hashing of
Broder \cite{broder97minwise,broder98minwise} is the estimation of
Jaccard set similarity $|A\cap B|/|A\cup B|$. Ignoring the probability of collisions in the minimum hash value,  we get
\begin{align*}
\Pr[ \min h(A) =& \min h(B) ] \\
&=~ \sum_{x\in A\cap B} \Pr[ h(x) = \min h(A\cup B)] 
\\
&=~ \frac{|A\cap B|}{|A\cup B|} \cdot \left( 1 \pm \widetilde{O}\left(
\frac{1}{|A\cup B|^{1/c}} \right) \right).
\end{align*}
For better bounds on the probabilities, we would make multiple
experiments with independent hash functions, yet this cannot eliminate
the bias $\eps$.

\paragraph{The power of two choices}
The power of two choices is a standard scheme for placing balls into
bins where each ball hashes to two bins, and is placed in the lightest
loaded one (see \cite{mitzenmacher01twochoice} for a survey). When placing $n$ balls
into $n$ bins, using the two-choice paradigm with truly random hash
functions, the maximum load of any bin is $\lg\lg n + O(1)$ w.h.p.
\cite{azar99lglgn}. Dahlgaard et
al. \cite{DKRT16:two-choice} have proved that simple tabulation gives
a maximum load which is $\lg\lg n + O(1)$ in expectation and $O(\log\log n)$ 
w.h.p.. This is the simplest constant time hashing scheme known to offer 
such strong two-choice load balancing.

\paragraph{Weakness with small numbers}
As described above, simple tabulation has much more power than
suggested by its 3-independence. However, there are also some
weaknesses. For example, in the Chernoff-style bounds \req{eq:char+}
and \req{eq:char-} from Theorem\tref{thm:chernoff}, we have an additive
error probability of $1/m^\gamma$ when hashing into $m$ bins. Here
$\gamma$ is an arbitrarily large constant, so this is fine when $m$ is
large. However, this is not good if $m$ is small, e.g., $m=2$ as when
we toss a coin. A somewhat related problem is that our bias $\eps$ in
$\eps$-minwise independence in \req{eq:char-min} is $\tilde O(1/n^{1/c})$ where $n$ is the
size of the set considered. This is fine if the set is large, but not
if the set is small. Both of these problems and more will be addressed
by twisted tabulation described below.

\section{Twisted tabulation}\label{sec:twisted}
We will now consider {\em twisted tabulation\/} proposed by
P\v{a}tra\c{s}cu and Thorup in \cite{PT13:twist}. It adds
a quick twist to simple tabulation,
leading to more general distributional properties, including Chernoff
bounds that also work for few bins and better minwise hashing that also works well for small sets.  For $i=1,\dots,c-1$, we expand the entries of
$h_i$ with a random character called the ``\emph{twister}''. More
precisely, for $i>0$, we now have random tables $h^\star_i:
\Sigma\rightarrow \Sigma\times\cR$.  This adds $\lg \Sigma$ bits to
each entry of the table (in practice, we want entries to have bit
lengths like 32 or 64, so depending on $\cR$, the twister may, or may
not, increase the actual entry size). The table $h_0:\Sigma\rightarrow
\cR$ is kept unchanged. The hash function is now computed in two
statements:
\begin{eqnarray}
  (t,h_{>0}) &=& \bigoplus_{i=1}^{c-1} h^\star_i[x_i]; \label{eq:twisted}
  \\
  h(x) &=& h_{>0} \oplus h_0[x_0 \oplus t]. \nonumber
\end{eqnarray}
Figure \ref{fig:code} contains the C-code for simple and twisted
tabulation.  
\begin{figure}
\begin{small}
\begin{verbatim}
#include <stdint.h> //defines uintX_t as unsigned X-bit integer. 

uint32_t SimpleTab32(uint32_t x, uint32_t[4][256] H) {
  uint32_t i;
  uint32_t h=0;
  uint8_t c;
  for (i=0;i<4;i++) {
    c=x;
    h^=H[i][c];
    x = x >> 8;
  }
  return h;
}

uint32_t TwistedTab32(uint32_t x, uint64_t[4][256] H) {
  uint32_t i;
  uint64_t h=0;
  uint8_t  c;
  for (i=0;i<3;i++) {
    c=x;
    h^=H[i][c];
    x = x >> 8;
  }                      
  c=x^h;                 // extra xor compared with simple
  h^=H[i][c];            
  h>>=32;                // extra shift compared with simple
  return ((uint32_t) h); 
}
\end{verbatim}
\end{small}
\caption{C-code for simple and twisted tabulation for
32-bit keys assuming a pointer H to some randomly filled storage
(4KB for simple and 8KB for twisted).}\label{fig:code}
\end{figure}

The highlight of twisted tabulation is its
minimalistic nature, adding very little to the cost of simple
tabulation (but, as we shall see, with significantly stronger
guarantees). Twisted tabulation uses exactly $c$ character lookups
into tables with $\Sigma$ entries, just like simple tabulation, though
with larger entries. Essentially twisted tabulation only differs from
simple tabulation by two AC$^0$ operations, so we would expect
it to be almost as fast as simple tabulation (whose practicality has
long been established). This was confirmed experimentally in \cite{PT13:twist}, where
twisted tabulation was less than 30\% slower than simple tabulation,
and still nearly three times faster than a second degree polynomial.

When we discuss properties of twisted tabulation, we view keys $x =
(x_0, \dots, x_{c-1})$ as composed of a head, $\head(x) = x_0$ and a
tail, $\tail(x) = (x_1, \dots, x_{c-1})$.  We refer to the following 
implementation of twisted tabulation which is less efficient but
mathematically equivalent to \eqref{eq:twisted}:
\begin{enumerate}
\item Pick a simple tabulation hash function $\twister : \Sigma^{c-1} \to
  \Sigma$ from $c-1$ characters to 1 character. This corresponds to
  the twister components of $h^\star_1, \dots, h^\star_{c-1}$. Applying $\twister$ 
   to the tail of a key $x$, 
   we get the combined twister $t=\twister(\tail(x))$, the ``twisted head''
   $x_0 \xor t$, and the ``twisted key''
   $\hT(x) = \big( x_0 \oplus t, x_1, \dots, x_{c-1}
  \big)$.

\item Pick a simple tabulation hash function $\hS : \Sigma^c \to \cR$
  (where $\cR$ was the desired output range). This corresponds to
  $h_0$ and for $i>0$, the non-twister component $h_i$ of $h^\star_i$
  (remember that tails are not touched by twisting). The twisted
  tabulation hash function is then $x \mapsto \hS(\hT(x))$.
\end{enumerate}

For all the results presented here, it does not
matter which character we view as the head. Above it is
the first character, but sometimes it is more efficient
if it is the last least significant character.

As noted in \cite{PT13:twist}, the twisting by $\hT$ can be seen as a
single-round Feistel permutation where the hash function is simple
tabulation.  Because twisting is a permutation, twisted tabulation
inherit from simple tabulation any (probabilistic) property that holds
regardless of concrete key values, e.g., the above Cuckoo hashing and
the power of two choices. Like simple tabulation, twisted tabulation is
only 3-independent, but it does have some more general distributional 
guarantees, which we explain in detail below. 

\paragraph{Chernoff Bounds}
Chernoff bounds play a prominent role in the design of randomized algorithms
\cite[\S 4]{motwani95book}. The Chernoff-style bounds from Theorem \ref{thm:chernoff} where limited in that they only really worked for throwing balls into
a large number $m$ of bins. In \cite{PT13:twist}, P\v{a}tra\c{s}cu and Thorup prove the following far more general Chernoff-style bounds for twisted tabulation.
\begin{theorem}[\cite{PT13:twist}]  \label{thm:chernoff-twist}
Choose a random $c$-character twisted tabulation hash function $h=\hS\circ\hT: [u]
\to [u]$, $[u]=\Sigma^c$. For each key
$x\in[u]$ in the universe, we have an arbitrary ``value function''
$v_x:[u] \to [0,1]$ assigning a value $V_x=v_x(h(x))\in [0,1]$ to $x$
for each hash 
value. Define $V = \sum_{x \in [u]} V_x$ and $\mu =\sum_{x \in [u]} \mu_x$ where $\mu_x=\E[v_x(h(x))]$.  Let
$\gamma$, $c$ and $\eps>0$ be constants. Then for any $\mu <
\Sigma^{1-\eps}$ and  $\delta>0$, we have:
\begin{align}\label{eq:twist+}
\Pr[V\ge (1+\delta)\mu] &\le \left(
   \frac{e^\delta}{(1+\delta)^{(1+\delta)}} 
             \right)^{\Omega\drop{_{\gamma,c, \eps}}(\mu)} + 
    1/u^\gamma\\
\Pr[V\le (1-\delta)\mu] &\le \left(
   \frac{e^{-\delta}} {(1-\delta)^{(1-\delta)}} \right)^{\Omega\drop{_{\gamma,
       c, \eps}}(\mu)} + 1/u^\gamma.\label{eq:twist-}
\end{align}
Moreover, for any $\mu\geq \sqrt \Sigma$ (including $\mu \geq 
\Sigma^{1-\eps}$), with probability $1-u^{-\gamma}$,
\begin{equation}\label{eq:large-mu}
V=\mu\pm \widetilde O\drop{_{\gamma,c}}(\sqrt{\mu}).
\end{equation}
If we have a given distinguished query key $q\in[u]$,
the above bounds hold even if we condition everything on $h(q)=a$ for
any given $a\in[u]$\footnote{This last statement conditioning on
$h(q)=a$ was not proved in \cite{PT13:twist}, but it is an easy
extension provided in Appendix \ref{sec:condition}.}.
\end{theorem}

The statement of Theorem \ref{thm:chernoff-twist} may seem a bit
cryptic, so before proceeding, we will show that it improves
the simple tabulation bounds from Theorem \ref{thm:chernoff}.
Those bounds considered a set $S$ of $n$ balls or keys mapped into $m$
bins, and had an error
bound of $m^{-\gamma}$. The error is here improved to
$u^{-\gamma}$.

We now do the translation. Discounting all irrelevant keys
$x\in[u]\setminus S$, we zero their value functions, setting
$v_x(\cdot)=0$. Also, we define the bin hash function from Theorem
\ref{thm:chernoff} as $h'(x)=h(x)\bmod m$, noting that $m\uparrow u$ since both are powers of two.
Theorem \ref{thm:chernoff} studies the number of balls landing a bin
$b$ which may be a function of the bin $h'(q)$ of a query ball 
$q\in[u]\setminus S$. Thanks to the last statement of Theorem
\ref{thm:chernoff-twist}, we can condition on any value $a$ of $h(q)$,
which determines $h'(q)$ and hence $b$. Now, for $x\in S$, define 
$V_x=v_x(h(x))=1$ if $h'(x)=b$; 0 otherwise. 
Now $V=\sum_{x\in[u]} V_x$ is the variable $X$ from Theorem \ref{thm:chernoff}, and
\req{eq:char+} and \req{eq:char-} follow from \req{eq:twist+} and
\req{eq:twist-}, but with the improved error $u^{-\gamma}$.
This is important when $m$ is small, e.g., if $m=2$ corresponding
to unbiased coin tosses.

A different illustration of the versatility of Theorem \ref{thm:chernoff-twist} is if we want each key $x\in [u]$ to be sampled with
some individual sampling probability $p_x$. In this case we have
no distinguished query key, and we can just
define $v_x(y)=1$ if $y<p_x\cdot m$; otherwise $v_x(y)=0$. Since
$h(x)$ is uniform in $[m]$, we have that $x$
  is sampled with $V_x=v_x(h(x))=1$ with probability $\tilde p_x=\lceil p_xm\rceil/m$. The
number $V=\sum_{x\in[u]} V_x$ of samples is now concentrated
according to \req{eq:twist+} and \req{eq:twist-}.

\paragraph{Minwise independence}
Concerning minwise hashing, Dahlgaard and Thorup \cite{DT14:twist-min} have 
proved that twisted tabulation yields the following strengthening of
Theorem \ref{thm:char-min} for simple tabulation.
\begin{theorem}[{\cite{DT14:twist-min}}]\label{thm:twist-min}
Consider a set $S\subseteq \Sigma^c$ of $n=|S|$ keys and $q\in S$. If
$h:\Sigma^c\rightarrow [m]$, $m\geq n u^{1/c}$ is implemented by twisted 
tabulation then:
\begin{equation}\label{eq:twist-min} \Pr[ h(q) = \min h(S)] = \frac{1\pm \eps}{n}, \qquad
\textrm{where } \eps = O\left( \frac{\lg^2 u}{u^{1/c}} \right). 
\end{equation}
\end{theorem}
The important difference is that the bias $O\left( \frac{\lg^2
  n}{n^{1/c}} \right)$ from Theorem \ref{thm:char-min} is replaced by 
$O\left(\frac{\lg^2 u}{u^{1/c}} \right)$ which is small regardless of the set
size. Such an absolutely small bias generally requires $\Omega(\log u)$-independence \cite{patrascu10kwise-lb}.

\paragraph{Short range amortization for chaining and linear probing}\label{ssec:amort}
We now switch to a quite different illustration of the power of
twisted tabulation hashing from \cite{PT13:twist}. Consider linear probing in a half-full hash table
with $n$ keys. Out of $\sqrt n$ operations, we expect some to take
$\Omega(\log n)$ time. Nevertheless we show that any window of $\log n$
operations on distinct keys is executed in $O(\lg n)$ time with high
probability.  This also holds for the simpler case of chaining.

The general point is that for any set of stored keys and any set of
window keys, the operation times within the window are sufficiently
independent that the average concentrates nicely around the expected
constant operation time. Such concentration of the average should not
be taken for granted with real hash functions.  In
\cite{patrascu10kwise-lb} are input examples for linear probing
with multiplication-shift hashing \cite{dietzfel96universal} such that
if one operation is slow, then most operations are slow. In  \cite{PT13:twist}
is presented a parallel universe construction causing similar problems for
simple tabulation.  As stated above, twisted tabulation does, however, provide
sufficient independence, and we expect this to prove useful in
other applications.

\paragraph{Pseudo-random numbers generators}
Like \emph{any} hash function, twisted tabulation naturally implies a
pseudo-random numbers generators (PRG) with the pseudo-random sequence
$h(0), h(1), \ldots$. For maximal efficiency, we use
the last and least significant character as the head. Thus,
a key $x=(x_{c-1},\ldots,x_0)\in\Sigma^c$ has
$\head(x)=x_0$ and $\tail(x)=x_{>0}=(x_{c-1},\ldots,x_1)$.
For twisted tabulation, we use a simple tabulation function $h^\star:\Sigma^{c-1}\to\Sigma\times [2^r]$
and a character function $h_0:\Sigma\to[2^r]$, and 
then $h(x)$ is computed, setting $(t,h_{>0})=h^\star(x_{>0})$
and $h(x)=h_0[x_0\xor t]\xor h_{>0}$.
We now think of $x$ as the pair $(x_{>0},x_0)\in \Sigma^{c-1}\times\Sigma$. As we increase the index
$x=0,1,\ldots,\Sigma-1,\Sigma,\Sigma+1,\ldots =(0,0),(0,1),\ldots,(0,\Sigma-1),(1,0),(1,1),\ldots$, the tail
$x_{>0}$ only increases when $x_0$ wraps around to $0$---once in every
$\Sigma$ calls.  We therefore store
$(t,h_{>0})=h^\star(x_{>0})$ in a register, recomputing it only when
$x_{>0}$ increases. Otherwise we compute $h(x)=h_0[x_0\xor
  t]\xor h_{>0}$ using just one character lookup and two $\xor$-operations. We found this to be exceedingly fast: as fast as
a single multiplication and 4 times faster than the standard random
number generator \texttt{random()} from the GNU C library which has almost no
probabilistic guarantees. Besides
being faster, the twisted PRG offers the powerful distributional
guarantees discussed above.

As an alternative implementation, we note that $h^\star$ is itself applied to
consecutive numbers $x_{>0}=0,1,2,\ldots$, so $h^\star$ can also be
implemented as a PRG. The $h^\star$-PRG is only applied once for every
$\Sigma$ numbers generated by $h$, so the $h^\star$-PRG can be much
slower without affecting the overall performance. Instead of
implementing $h^\star$ by simple tabulation, we could implement it
with any logarithmically independent PRG, thus not storing any tables
for $h^\star$, but instead generating each new value $h^\star(x_{>0})$
on the fly as $x_{>0}$ increases. We can view this as a general 
conversion of a comparatively slow but powerful PRG into an extremely fast one preserving
the many probabilistic properties of twisted tabulation. The
recent PRG of Christiani and
Pagh \cite{christiani14prg} would be a good choice for the
$h^\star$-PRG if we do not want to implement it with simple tabulation.

\paragraph{Randomized Algorithms and Data Structures}
When using the twisted PRG in randomized algorithms \cite{motwani95book},
we get the obvious advantage of the Chernoff-style bounds from
Theorem \ref{thm:chernoff-twist} which is one of the basic
techniques needed \cite[\S 4]{motwani95book}.
The $\eps$-minwise hashing from Theorem \ref{thm:twist-min} with 
$\eps=\tO(1/u^{1/c})$ is important in contexts where we
want to assign randomized priorities to items.  A direct example is
treaps \cite{seidel96treaps}. The classic analysis \cite[\S
  8.2]{motwani95book} of the expected operation time is based on each
key in an interval having the same chance of getting the lowest
priority. Assigning the priorities with an $\eps$-minwise hash
function, expected cost is only increased by a factor $(1+\eps)$ compared with
the unrealistic case of true randomness. In static
settings, we can also use this to generate a random permutation,
sorting items according to priorities. This sorting itself takes
linear time since we essentially only need to look at the $\log n$
most significant bits to sort $n$ priorities. Using this order to pick
items, we get that classic algorithms like QuickSort \cite[\S 1]{motwani95book} and Binary Planar
Partitions \cite[\S 1.3]{motwani95book} perform within an expected factor
$(1+\eps)$ of what they would with true randomness. With
$\eps=\tO(1/u^{1/c})$ as with our twisted PRG, this is very close
the expected performance with true randomness.

\section{Double tabulation and high independence}
Thorup \cite{Tho13:simple-simple} has shown that simple tabulation can also
be used to get high independence if we apply it twice. More precisely,
we consider having two independent simple tabulation functions $h_0:\Sigma^c\fct\Sigma^d$ and $h_1:\Sigma^d\fct[2^r]$, and then the claim is that
$h_1\circ h_0$ is likely to be highly independent. The main point from \cite{Tho13:simple-simple} is that
the first simple tabulation $h_0$ is likely to have an expander-type
property. 

More precisely, given a function $f:[u]\fct\Sigma^d$,
a key set $X\subseteq [u]$ has a {\em unique output
character\/} if there is a key $x\in X$ and a $j\in[d]$ and such that for
all $y\in X\setminus\{x\}$, $h(y)_j\neq h(x)_j$, that is, the $j$th
output character is unique to some $x$ in $X$. 
We say that $f$ is {\em $k$-unique\/} if each non-empty key set 
$Y\subseteq [u]$ of size at most $k$ has a unique output character.
Siegel \cite{siegel04hash} noted that
if $f$ is $k$-unique and 
$h_1:\Sigma^d\fct[2^r]$ is a random simple tabulation function,
then $h_1\circ f:[u]\fct[2^r]$ is $k$-independent.
The main technical result from \cite{Tho13:simple-simple} is
\begin{theorem}[{\cite{Tho13:simple-simple}}] \label{thm:double-thorup}
Consider a random simple tabulation function $h_0:\Sigma^c\fct \Sigma^d$.
Assume $c=\Sigma^{o(1)}$ and $(c+d)^c=\Sigma^{o(1)}$.  Let
$k=\Sigma^{1/(5c)}$. With probability
$1-o(\Sigma^{2-d/(2c)})$, the function $h_0$ is $k$-unique. 
More concretely for 32-bit keys with 16-bit characters, $h_0$ is 100-unique
with probability $1-1.5\times 10^{-42}$.

Assuming that $h_0$ is $k$-unique, if $h_1:\Sigma^d\fct[2^r]$ is
a random simple tabulation function, then $h_1\circ h_0$ is $k$-independent.
\end{theorem}
This construction for highly independent hashing is much simpler than that of 
Siegel \cite{siegel04hash} mentioned in Section\tref{sec:intro},
and for $d=O(c)$, the evaluation takes $O(c)$ time as opposed to
the $O(c)^c$ time used by Siegel.

Complementing the above result, Dahlgaard et al. \cite{DKRT15:k-part} have
proved that double tabulation is likely to be truly random for
any specific set $S$ with less than $(1-\Omega(1))\Sigma$ keys:
\begin{theorem}[{\cite{DKRT15:k-part}}]\label{thm:double-uniform}
Given a set $S\subseteq [u]$ of size $(1-\Omega(1))\Sigma$, consider
two random simple tabulation function $h_0:\Sigma^c\fct \Sigma^d$ and
$h_1:\Sigma^d\fct [2^r]$.  With probability $1-O(\Sigma^{1-\lfloor
  d/2\rfloor})$, every non-empty subset $X\subseteq S$ gets a unique
output character with $h_0$, and then the double tabulation function
$h_1\circ h_0$ is fully random over $S$.
\end{theorem}
It is interesting to compare Theorem\tref{thm:double-uniform} and
Theorem\tref{thm:double-thorup}. Theorem\tref{thm:double-uniform} holds for one large set while  Theorem\tref{thm:double-thorup} works for all small sets. Also,
Theorem\tref{thm:double-uniform} with $d=4$ ``derived'' characters gets essentially
the same error probability as Theorem\tref{thm:double-thorup} with $d=6c$ derived characters.

Siegel \cite{siegel04hash} has proved that
with space $\Sigma$, we cannot in constant time hope to get independence higher
than $\Sigma^{1-\Omega(1)}$, which is much less than
the size of the given set in Theorem \ref{thm:double-uniform}. 

Theorem\tref{thm:double-uniform} provides an extremely simple $O(n)$
space implementation of a constant time hash function that is likely
uniform on any given set $S$.  This should be compared with the
previous linear space uniform hashing of Pagh and Pagh \cite[\S
  3]{PP08}.  The most efficient implementation of \cite[\S 3]{PP08}
uses the highly independent double tabulation from
Theorem\tref{thm:double-thorup} a subroutine. However, as discussed
earlier, double tabulation uses much more derived characters for high
independence than for uniformity on a given set, so for linear space
uniform hashing on a given set, it is much faster and simpler to use
the double tabulation of Theorem\tref{thm:double-uniform} directly. We note that \cite[\S 4]{PP08} presents
a general trick to reduce the space from linear, that is, $O(n(\lg
n+\lg|\cR|))$ bits, downto $(1+\eps)n\lg|\cR| + O(n)$ bits,
preserving the constant evaluation time. This reduction can also be
applied to Theorem\tref{thm:double-uniform} so that we also get
a simpler overall construction for a succinct dictionary
using $(1+\eps)n\lg|\cR| + O(n)$ bits of space and constant evaluation time.

Very recently, Christiani et al. \cite{CPT15:indep} have shown that
we using a more elaborate recursive tabulation scheme can get quite to Siegel's lower-bound.
\begin{theorem}[{\cite[Corollary 3]{CPT15:indep}}]\label{thm:high-ind}
For word-size $w$, and parameters $k$ and $c=O(w/(\log k))$,
with probability $1-u^{1/c}$, we can construct a $k$-independent
hash function $h:[2^w]\fct[2^w]$ in $O(cku^{1/c})$ time and space that
is evaluated in $O(c\log c)$ time.
\end{theorem}
In Theorem \ref{thm:double-uniform}, we used the same space to get 
independence $\Sigma^{1/(5c)}$ and evaluation time $O(c)$, and
the construction of Theorem \ref{thm:double-uniform} is simpler. We should thus use Theorem \ref{thm:high-ind} if we
need its very high independence, but if, say, logarithmic
independence suffices, then Theorem \ref{thm:double-uniform} is
the better choice. 

A major open problem is get the space and independence
of Theorem \ref{thm:high-ind} but with $O(c)$ evaluation time, matching
the lower bound of \cite{siegel04hash}. In its full generality,
the lower bound from \cite{siegel04hash} says that we for independence
$k$ with $c<k$ cell probes need space $\Omega(k(u/k)^{1/c}c)$.

\paragraph{Invertible Bloom filters with simple tabulation}
Theorem \ref{thm:double-uniform} states that if a 
random simple tabulation function $h_0:\Sigma^c\fct \Sigma^d$ 
is applied to a given set $S$ of size $(1-\Omega(1))\Sigma$,
then with probability $1-O(\Sigma^{1-\lfloor
  d/2\rfloor})$, every non-empty subset $X\subseteq S$ gets a unique
output character. This is not only relevant for fully-random hashing. This property is
also sufficient for the hash function in Goodrich and Mitzenmacher's Invertible
Bloom Filters \cite{Goodrich11ibt}, which have found numerous applications in
streaming and data bases
\cite{eppstein2011straggler,eppstein2011s,mitzenmacher2012biff}.
So far Invertible Bloom Filters have been implemented
with fully random hashing, but Theorem \ref{thm:double-uniform} shows that simple
tabulation suffices for the underlying hash function.

\subsection{$k$-partitions via mixed tabulation}
The general goal of Dahlgaard et al. \cite{DKRT15:k-part} is
a hash function for $k$-partitioning
a set into bins so that we get good concentration bounds when combining
statistics from each bin.

To understand this point, suppose we have a
fully random hash function applied to a set $X$  of red and blue balls.
We want to estimate the fraction $f$ of red balls. The idea of Minwise hashing
is to sample the ball with the smallest hash value. This sample is uniformly
random and is red with probability $f$. If we repeat the experiment
$k$ times with $k$ independent hash functions, we get
a multiset $S$ of $k$ samples with replacement from $X$ and the
fraction red balls in $S$ concentrates around $f$ as we increase the number
of samples.

Consider the alternative experiment using a single hash function, where
we use some bits of the hash value to partition $X$ into $k$ bins, and
then use the remaining bits as a local hash value. We
pick the ball with the smallest hash value in each bin. This is a sample $S$
from $X$ without replacement, and again, the fraction of red balls
is concentrated around $f$.

The big difference between the two schemes is that the second one runs
$\Omega(k)$ times faster. In the first experiment, each ball
participated in $k$ independent experiments, but in the second one with
$k$-partitions, each ball picks its bin, and then only
participates in the local experiment for that bin. Thus with the $k$-partition, essentially, we get
$k$ experiments for the price of one.

This generic idea has been used for different types of statistics.
Flajolet and Martin \cite{Flajolet85counting} introduced it to count the number of
distinct items in a multiset, Charikar et al.~~\cite{Charikar02countsketch} 
used it in their count
sketches for fast estimation of the second moment of a data stream,
and recently,~Li et
al.~\cite{li12oneperm,Shrivastava14oneperm} used it for Minwise estimation of the Jaccard Similarity of two sets.

The issue is that no realistic hashing scheme was known to make a good
enough $k$-partition for the above kind of statistics to make sense.
The point is that the contents of different
bins may be too correlated, and then we get no better
concentration with a larger $k$. In the independence paradigm of 
Carter and Wegman~\cite{wegman81kwise}, it would seem that we need
independence at least $k$ to get sufficiently independent statistics
from the different bins. 

An efficient solution is based on a variant of double tabulation described
below.

\paragraph{Mixed tabulation} For Theorem~\ref{thm:double-uniform}
we may use $d=4$ even if $c$ is larger, but then $h_0$ will introduce
many collisions. To avoid this problem we mix the schemes in \emph{mixed
tabulation}. Mathematically, we use two simple tabulation hash functions
$h_1 : [u]\to \Sigma^d$ and $h_2 : \Sigma^{c+d}\to [2^r]$, and define the hash
function $h(x)\mapsto h_2(x\circ h_1(x))$, where $\circ$ denotes
concatenation of characters. We call $x\circ h_1(x)$ the \emph{derived
key}, consisting of $c$ original characters and $d$ derived characters. Since the derived keys includes the original keys, there are no duplicate
keys. 

We note that mixed tabulation only requires $c+d$ lookups if we
instead store simple tabulation functions $h_{1,2}:\Sigma^c\to
\Sigma^d\times [r]$ and $h_2':\Sigma^d\to [r]$, computing $h(x)$ by
$(v_1,v_2)=h_{1,2}(x);\ h(x)=v_1\xor h_2(v_2)$.  This efficient
implementation is similar to that of twisted tabulation, and is
equivalent to the previous definition. As long as we have at least one
derived character, mixed tabulation has all the distribution
properties of twisted tabulation, particularly, the Chernoff-style
concentration bound from Theorem \ref{thm:chernoff-twist}. At the
same time, we get the full randomness from Theorem \ref{thm:double-uniform}
for any given set $S$ of size $(1-\Omega(1))\Sigma$. Based on these properties
and more, it is proved in \cite{DKRT15:k-part} that mixed tabulation, w.h.p.,
gets essentially the same concentration bounds as full randomness for all of the above mentioned statistics based on $k$-partitions.


\begin{thebibliography}{10}

\bibitem{alon99linear}
N.~Alon, M.~Dietzfelbinger, P.~B. Miltersen, E.~Petrank, and G.~Tardos.
\newblock Linear hash functions.
\newblock {\em J. {ACM}}, 46(5):667--683, 1999.

\bibitem{alon96frequency}
N.~Alon, Y.~Matias, and M.~Szegedy.
\newblock The space complexity of approximating the frequency moments.
\newblock {\em Journal of Computer and System Sciences}, 58(1):209--223, 1999.
\newblock Announced at STOC'96.

\bibitem{alon08kwise}
N.~Alon and A.~Nussboim.
\newblock {$k$}-wise independent random graphs.
\newblock In {\em Proc. 49th IEEE Symposium on Foundations of Computer Science
  (FOCS)}, pages 813--822, 2008.

\bibitem{azar99lglgn}
Y.~Azar, A.~Z. Broder, A.~R. Karlin, and E.~Upfal.
\newblock Balanced allocations.
\newblock {\em SIAM Journal on Computing}, 29(1):180--200, 1999.
\newblock Announced at STOC'94.

\bibitem{black98linprobe}
J.~R. Black, C.~U. Martel, and H.~Qi.
\newblock Graph and hashing algorithms for modern architectures: Design and
  performance.
\newblock In {\em Proc. 2nd International Workshop on Algorithm Engineering
  (WAE)}, pages 37--48, 1998.

\bibitem{braverman10kwise}
V.~Braverman, K.-M. Chung, Z.~Liu, M.~Mitzenmacher, and R.~Ostrovsky.
\newblock {AMS} without 4-wise independence on product domains.
\newblock In {\em Proc. 27th Symposium on Theoretical Aspects of Computer
  Science (STACS)}, pages 119--130, 2010.

\bibitem{broder98minwise}
A.~Z. Broder, M.~Charikar, A.~M. Frieze, and M.~Mitzenmacher.
\newblock Min-wise independent permutations.
\newblock {\em Journal of Computer and System Sciences}, 60(3):630--659, 2000.
\newblock Announced at STOC'98.

\bibitem{broder97minwise}
A.~Z. Broder, S.~C. Glassman, M.~S. Manasse, and G.~Zweig.
\newblock Syntactic clustering of the web.
\newblock {\em Computer Networks}, 29:1157--1166, 1997.

\bibitem{carter77universal}
L.~Carter and M.~N. Wegman.
\newblock Universal classes of hash functions.
\newblock {\em Journal of Computer and System Sciences}, 18(2):143--154, 1979.
\newblock Announced at STOC'77.

\bibitem{Charikar02countsketch}
M.~Charikar, K.~Chen, and M.~Farach-Colton.
\newblock Finding frequent items in data streams.
\newblock In {\em Proc. 29th International Colloquium on Automata, Languages
  and Programming (ICALP)}, pages 693--703. Springer-Verlag, 2002.

\bibitem{christiani14prg}
T.~Christiani and R.~Pagh.
\newblock Generating k-independent variables in constant time.
\newblock In {\em Proceedings of the 55th IEEE Symposium on Foundations of
  Computer Science (FOCS)}, pages 196--205, 2014.

\bibitem{CPT15:indep}
T.~Christiani, R.~Pagh, and M.~Thorup.
\newblock From independence to expansion and back again.
\newblock In {\em Proceedings of the 47th {ACM} Symposium on Theory of
  Computing ({STOC})}, pages 813--820, 2015.

\bibitem{DKRT15:k-part}
S.~Dahlgaard, M.~B.~T. Knudsen, E.~Rotenberg, and M.~Thorup.
\newblock Hashing for statistics over k-partitions.
\newblock In {\em Proceedings of the 56th IEEE Symposium on Foundations of
  Computer Science (FOCS)}, pages 1292--1310, 2015.

\bibitem{DKRT16:two-choice}
S.~Dahlgaard, M.~B.~T. Knudsen, E.~Rotenberg, and M.~Thorup.
\newblock The power of two choices with simple tabulation.
\newblock In {\em Proceedings of the 27th {ACM-SIAM} Symposium on Discrete
  Algorithms (SODA)}, pages 1631--1642, 2016.

\bibitem{DT14:twist-min}
S.~Dahlgaard and M.~Thorup.
\newblock Approximately minwise independence with twisted tabulation.
\newblock In {\em Proc. 14th Scandinavian Workshop on Algorithm Theory (SWAT)},
  pages 134--145, 2014.

\bibitem{dietzfel96universal}
M.~Dietzfelbinger.
\newblock Universal hashing and $k$-wise independent random variables via
  integer arithmetic without primes.
\newblock In {\em Proc. 13th Symposium on Theoretical Aspects of Computer
  Science (STACS)}, pages 569--580, 1996.

\bibitem{dietzfel97closest}
M.~Dietzfelbinger, T.~Hagerup, J.~Katajainen, and M.~Penttonen.
\newblock A reliable randomized algorithm for the closest-pair problem.
\newblock {\em Journal of Algorithms}, 25(1):19--51, 1997.

\bibitem{dietzfel90highperf}
M.~Dietzfelbinger and F.~{Meyer auf der Heide}.
\newblock A new universal class of hash functions and dynamic hashing in real
  time.
\newblock In {\em Proc. 17th International Colloquium on Automata, Languages
  and Programming (ICALP)}, pages 6--19, 1990.

\bibitem{dietzfelbinger09cuckoo-bas}
M.~Dietzfelbinger and U.~Schellbach.
\newblock On risks of using cuckoo hashing with simple universal hash classes.
\newblock In {\em Proc. 20th ACM/SIAM Symposium on Discrete Algorithms (SODA)},
  pages 795--804, 2009.

\bibitem{dietzfel03tabhash}
M.~Dietzfelbinger and P.~Woelfel.
\newblock Almost random graphs with simple hash functions.
\newblock In {\em Proc. 25th ACM Symposium on Theory of Computing (STOC)},
  pages 629--638, 2003.

\bibitem{eppstein2011straggler}
D.~Eppstein and M.~T. Goodrich.
\newblock Straggler identification in round-trip data streams via newton's
  identities and invertible bloom filters.
\newblock {\em Knowledge and Data Engineering, IEEE Transactions on},
  23(2):297--306, 2011.

\bibitem{eppstein2011s}
D.~Eppstein, M.~T. Goodrich, F.~Uyeda, and G.~Varghese.
\newblock What's the difference?: efficient set reconciliation without prior
  context.
\newblock {\em ACM SIGCOMM Computer Communication Review}, 41(4):218--229,
  2011.

\bibitem{Flajolet85counting}
P.~Flajolet and G.~N. Martin.
\newblock Probabilistic counting algorithms for data base applications.
\newblock {\em Journal of Computer and System Sciences}, 31(2):182--209, 1985.

\bibitem{fredman84dict}
M.~L. Fredman, J.~Koml\'{o}s, and E.~Szemer\'{e}di.
\newblock Storing a sparse table with {0(1)} worst case access time.
\newblock {\em Journal of the ACM}, 31(3):538--544, 1984.
\newblock Announced at FOCS'82.

\bibitem{Goodrich11ibt}
M.~T. Goodrich and M.~Mitzenmacher.
\newblock Invertible {b}loom lookup tables.
\newblock In {\em Proceedings of the 49th Allerton Conference on Communication,
  Control, and Computing}, pages 792--799, 2011.

\bibitem{heileman05linprobe}
G.~L. Heileman and W.~Luo.
\newblock How caching affects hashing.
\newblock In {\em Proc. 7th Workshop on Algorithm Engineering and Experiments
  (ALENEX)}, pages 141--–154, 2005.

\bibitem{indyk01minwise}
P.~Indyk.
\newblock A small approximately min-wise independent family of hash functions.
\newblock {\em Journal of Algorithms}, 38(1):84--90, 2001.
\newblock Announced at SODA'99.

\bibitem{cohen09cuckoo5}
D.~M.~K. Jeffery S.~Cohen.
\newblock Bounds on the independence required for cuckoo hashing, 2009.
\newblock Manuscript.

\bibitem{karloff93prg}
H.~J. Karloff and P.~Raghavan.
\newblock Randomized algorithms and pseudorandom numbers.
\newblock {\em Journal of the ACM}, 40(3):454--476, 1993.

\bibitem{knuth63linprobe}
D.~E. Knuth.
\newblock Notes on open addressing.
\newblock Unpublished memorandum. See
  \texttt{http://citeseer.ist.psu.edu/knuth63notes.html}, 1963.

\bibitem{knuth-vol3}
D.~E. Knuth.
\newblock {\em The Art of Computer Programming, Volume {III}: {S}orting and
  Searching}.
\newblock Addison-Wesley, 1973.

\bibitem{li12oneperm}
P.~Li, A.~B. Owen, and C.-H. Zhang.
\newblock One permutation hashing.
\newblock In {\em Proc. 26th Conference on Neural Information Processing
  Systems (NIPS)}, pages 3122--3130, 2012.

\bibitem{mitzenmacher01twochoice}
M.~Mitzenmacher, A.~W. Richa, and R.~Sitaraman.
\newblock The power of two random choices: A survey of techniques and results.
\newblock In P.~Pardalos, S.~Rajasekaran, and J.~Rolim, editors, {\em Handbook
  of Randomized Computing: volume 1}, pages 255--312. Springer, 2001.

\bibitem{mitzenmacher08hash}
M.~Mitzenmacher and S.~P. Vadhan.
\newblock Why simple hash functions work: exploiting the entropy in a data
  stream.
\newblock In {\em Proc. 19th ACM/SIAM Symposium on Discrete Algorithms (SODA)},
  pages 746--755, 2008.

\bibitem{mitzenmacher2012biff}
M.~Mitzenmacher and G.~Varghese.
\newblock Biff (bloom filter) codes: Fast error correction for large data sets.
\newblock In {\em Information Theory Proceedings (ISIT), 2012 IEEE
  International Symposium on}, pages 483--487. IEEE, 2012.

\bibitem{motwani95book}
R.~Motwani and P.~Raghavan.
\newblock {\em Randomized Algorithms}.
\newblock Cambridge University Press, 1995.

\bibitem{PP08}
A.~Pagh and R.~Pagh.
\newblock Uniform hashing in constant time and optimal space.
\newblock {\em SIAM Journal on Computing}, 38(1):85--96, 2008.

\bibitem{pagh07linprobe}
A.~Pagh, R.~Pagh, and M.~Ru{\v z}i{\'c}.
\newblock Linear probing with constant independence.
\newblock {\em SIAM Journal on Computing}, 39(3):1107--1120, 2009.
\newblock Announced at STOC'07.

\bibitem{pagh04cuckoo}
R.~Pagh and F.~F. Rodler.
\newblock Cuckoo hashing.
\newblock {\em Journal of Algorithms}, 51(2):122--144, 2004.
\newblock Announced at ESA'01.

\bibitem{patrascu10kwise-lb}
M.~P{\v a}tra{\c s}cu and M.~Thorup.
\newblock On the $k$-independence required by linear probing and minwise
  independence.
\newblock In {\em Proc. 37th International Colloquium on Automata, Languages
  and Programming (ICALP)}, pages 715--726, 2010.

\bibitem{patrascu12charhash}
M.~P{\v a}tra{\c s}cu and M.~Thorup.
\newblock The power of simple tabulation-based hashing.
\newblock {\em Journal of the ACM}, 59(3):Article 14, 2012.
\newblock Announced at STOC'11.

\bibitem{PT13:twist}
M.~P\v{a}tra\c{s}cu and M.~Thorup.
\newblock Twisted tabulation hashing.
\newblock In {\em Proc. 24th ACM/SIAM Symposium on Discrete Algorithms (SODA)},
  pages 209--228, 2013.

\bibitem{schmidt95chernoff}
J.~P. Schmidt, A.~Siegel, and A.~Srinivasan.
\newblock Chernoff-{H}oeffding bounds for applications with limited
  independence.
\newblock {\em SIAM Journal on Discrete Mathematics}, 8(2):223--250, 1995.
\newblock Announced at SODA'93.

\bibitem{seidel96treaps}
R.~Seidel and C.~R. Aragon.
\newblock Randomized search trees.
\newblock {\em Algorithmica}, 16(4/5):464--497, 1996.
\newblock Announced at FOCS'89.

\bibitem{Shrivastava14oneperm}
A.~Shrivastava and P.~Li.
\newblock Densifying one permutation hashing via rotation for fast near ne\
  ighbor search.
\newblock In {\em Proc. 31th International Conference on Machine Learning
  (ICML)}, pages 557--565, 2014.

\bibitem{siegel04hash}
A.~Siegel.
\newblock On universal classes of extremely random constant-time hash
  functions.
\newblock {\em SIAM Journal on Computing}, 33(3):505--543, 2004.
\newblock Announced at FOCS'89.

\bibitem{thorup00universal}
M.~Thorup.
\newblock Even strongly universal hashing is pretty fast.
\newblock In {\em Proc. 11th ACM/SIAM Symposium on Discrete Algorithms (SODA)},
  pages 496--497, 2000.

\bibitem{thorup09linprobe}
M.~Thorup.
\newblock String hashing for linear probing.
\newblock In {\em Proc. 20th ACM/SIAM Symposium on Discrete Algorithms (SODA)},
  pages 655--664, 2009.

\bibitem{thorup11timerev}
M.~Thorup.
\newblock Timeouts with time-reversed linear probing.
\newblock In {\em Proc. IEEE INFOCOM}, pages 166--170, 2011.

\bibitem{Tho13:simple-simple}
M.~Thorup.
\newblock Simple tabulation, fast expanders, double tabulation, and high
  independence.
\newblock In {\em Proc. 54th IEEE Symposium on Foundations of Computer Science
  (FOCS)}, pages 90--99, 2013.

\bibitem{thorup12kwise}
M.~Thorup and Y.~Zhang.
\newblock Tabulation-based 5-independent hashing with applications to linear
  probing and second moment estimation.
\newblock {\em SIAM Journal on Computing}, 41(2):293--331, 2012.
\newblock Announced at SODA'04 and ALENEX'10.

\bibitem{wegman81kwise}
M.~N. Wegman and L.~Carter.
\newblock New classes and applications of hash functions.
\newblock {\em Journal of Computer and System Sciences}, 22(3):265--279, 1981.
\newblock Announced at FOCS'79.

\bibitem{zobrist70hashing}
A.~L. Zobrist.
\newblock A new hashing method with application for game playing.
\newblock Technical Report~88, Computer Sciences Department, University of
  Wisconsin, Madison, Wisconsin, 1970.

\end{thebibliography}

\appendix
\section{Conditioning on the hash value of a query key}\label{sec:condition}
We are here going to argue that the Chernoff-style bounds from 
Theorem \ref{thm:chernoff-twist} hold when we for a given query
key $q$ and hash value $a$ condition everything on $h(q)=a$.
This will be a simple extension of the proof of Theorem
\ref{thm:chernoff-twist} in \cite[\S 3]{PT13:twist} without the condition. Below we first describe
the essential points from the proof in \cite[\S 3]{PT13:twist}, referring the
reader to \cite[\S 3]{PT13:twist} for more details. Afterwards we describe 
the extension to handle the condition.
\paragraph{No conditioning.}
In \cite[\S 3]{PT13:twist}, the twisted tabulation hash function is picked 
in 3 steps.
\begin{enumerate}
\item\label{st:head-twist} We randomly fix $\hT$ twisting the keys. This
  defines the twisted groups $G_\alpha$
  consisting of keys with the same twisted head $\alpha$.
\item\label{st:tail-hash}  
We randomly fix the $\hS_i$, $i>0$,
hashing the tails of the keys.  
\item\label{st:Chernoff}  We randomly fix $\hS_0$, hashing the twisted heads. 
This finalizes the twisted tabulation hashing $h=\hS\circ\hT$.
\end{enumerate}
Let $V_\alpha=\sum_{x\in G_\alpha} v_x(h(x))$ be the total value of
keys in twisted group $G_\alpha$. The main technical point in \cite[\S
  3]{PT13:twist} is to prove that, w.h.p., after step
\ref{st:tail-hash}, we will have $V_\alpha\leq d=O(1)$ no matter how
we fix $\hS_0[\alpha]$, and this holds simultaneously for all twisted
groups. While $\hS_0[\alpha]$ has not been fixed, $h(x)$ is
uniform in $[u]$ for every $x\in G_\alpha$, so the expected value
$\mu_\alpha=\E[V_\alpha]=\sum_{x\in G_\alpha} \mu_x$ over $G_\alpha$
is unchanged before step \ref{st:Chernoff}.  It also follows from \cite[\S
  3]{PT13:twist} that $\mu_\alpha\leq d$.
Our final value is
$V=\sum_{\alpha\in \Sigma} V_\alpha$ which still has the correct mean
$\E[V]=\mu$. Moreover, $V$ is Chernoff-concentrated since it is summing
$d$-bounded variables $V_\alpha$ as we fix the $\hS_0[\alpha]$ in step
\ref{st:Chernoff}. We are here ignoring some technicalities explained
in \cite{patrascu12charhash,PT13:twist}, e.g., how to formally handle
if some variable $V_\alpha$ is not $d$-bounded.

\paragraph{Conditioning on \boldmath{$h(q)=a$}.}
We now consider the effects of the condition $h(q)=a$. First we 
note that since twisted tabulation is 2-independent, the
condition $h(q)=a$ does not affect the expected value
$\mu_x=\E[v_x(h(x))]$ of any key $x\neq q$.

Now, for the above step-wise fixing of the twisted tabulation hash
function, we note that the only effect of the condition $h(q)=a$ is in step \ref{st:Chernoff}.  If $\alpha_0$ is
the twisted head of the query key, then instead of picking
$\hS_0[\alpha_0]$ at random, we have to fix it as
$\hS_0[\alpha_0]=a\xor\bigoplus_{i=1}^{c-1}\hS_i[q_i]$. Since the
first steps were not affected, w.h.p, for all
$\alpha\in\Sigma$, we still have that $V_\alpha\leq d=O(1)$ no matter how we fix
$\hS_0[\alpha]$, and this includes the twisted head $\alpha_0$.

We are now again considering $V=\sum_{\alpha\in \Sigma} V_\alpha$ where
each $V_\alpha$ is $d$-bounded. For all $\alpha\neq \alpha_0$, we
have $\E[V_\alpha]=\mu_\alpha$. However, $V_{\alpha_0}$ is fixed 
to some value in $[0,d]$ when we after 
steps \ref{st:head-twist}--\ref{st:tail-hash} are forced to
set $\hS_0[\alpha_0]=a\xor\bigoplus_{i=1}^{c-1}\hS_i[q_i]$. 
The error $|V_{\alpha_0}-\mu_{\alpha_0}|$ from this fixing is less than $d$,
and this has no effect on our error probability bounds except in cases
where Markov's bound takes over. However, 
Markov's bound holds directly for twisted tabulation
without any of the above analysis. Thus we conclude that Theorem \ref{thm:chernoff-twist} holds also when we condition on $h(q)=a$.

\end{document}